\documentclass[pre,amsmath,aps,superscriptaddress,twocolumn,floatfix]{revtex4}
\usepackage{graphicx,xspace}
\usepackage{float}
\usepackage{amsmath}
\usepackage{amssymb}
\usepackage{color}
\usepackage[normalem]{ulem}
\input{epsf}

\begin{document}
\newcommand{\scalar}[2]{\left \langle#1\ #2\right \rangle}
\newcommand{\me}{\mathrm{e}}
\newcommand{\mi}{\mathrm{i}}
\newcommand{\dif}{\mathrm{d}}
\newcommand{\period}{\text{per}}
\newcommand{\free}{\text{fr}}
\newcommand{\eq}[1]{Eq.~(\ref{e:#1})}
\newcommand{\eqq}[1]{Eq.~(\ref{e:#1})}
\newcommand{\EQ}[1]{(\ref{e:#1})}
\newcommand{\eqtwo}[2]{equations~(\ref{e:#1}) and~(\ref{e:#2})}
\newcommand{\EQTWO}[2]{Equations~(\ref{e:#1}) and~(\ref{e:#2})}
\newcommand{\fig}[1]{Fig.~\ref{f:#1}}
\newcommand{\FIG}[1]{Fig.~\ref{f:#1}}
\newcommand{\quot}[1]{\lq#1\rq}
\newcommand{\eg}{\textrm{e.g.}}
\newcommand{\cf}{\textrm{cf}}
\newcommand{\etc}{\textrm{etc}}
\newcommand{\ie}{\textrm{i.e.}}
\newcommand{\SET}[1]{\{#1\}}
\newcommand{\expl}[1]{\exp \left[ #1 \right] } 
\newcommand{\lb}{\left[}  
\newcommand{\rb}{\right]}  
\newcommand{\lc}{\left(}  
\newcommand{\rc}{\right)}  
\newcommand{\mult}{\times} 
\newcommand{\multcc}{\cdot} 
\newcommand{\multcn}{\cdot} 
\newcommand{\multnn}{\cdot} 
\newcommand{\ran}{\sub{ran}}
\newcommand{\dd}[1]{\text{d}{#1\ }}   
\newcommand{\ddd}[1]{\text{d}{#1}}   
\newcommand{\scal}[2]{(#1 \pmb{\cdot} #2)}
\newcommand{\mean}[1]{\overline{#1}}
\newcommand{\half}{\frac{1}{2}}
\newcommand{\Eesc}[1]{E^{\text{esc}}_{#1}}
\newcommand{\Lopt}{L_{\text{opt}}}
\newcommand{\dep}{_\text{dep}}
\newcommand{\equi}{_\text{eq}}
\newcommand{\flow}{_\text{flow}}
\newcommand{\equil}{_\text{eq}}
\newcommand{\Lrelax}{L_{\text{relax}}}
\newcommand{\Ltrans}{L_{\text{trans}}}
\newcommand{\Lav}{L_{\text{av}}}
\newcommand{\therm}{\text{therm}}
\newcommand{\anh}{\text{anh}}
\newcommand{\esc}{\text{esc}}
\newcommand{\cut}{\text{cut}}
\newcommand{\chap}[1]{Chapter~\ref{c:#1}}
\newcommand{\sect}[1]{Section~\ref{s:#1}}
\newcommand{\SECT}[1]{\ref{s:#1}}
\newcommand{\app}[1]{Appendix~\ref{a:#1}}
\newcommand{\subsect}[1]{Subsection~\ref{s:#1}}
\newcommand{\SUBSECT}[1]{\ref{s:#1}}
\newcommand{\clabel}[1]{\label{c:#1}}
\newcommand{\elabel}[1]{\label{e:#1}}
\newcommand{\exlabel}[1]{\label{ex:#1}} 
\newcommand{\flabel}[1]{\label{f:#1}}
\newcommand{\slabel}[1]{\label{s:#1}}
\newcommand{\alabel}[1]{\label{a:#1}}

\title{Creep dynamics of elastic manifolds \emph{via} exact transition pathways}

\author{Alejandro B.
Kolton}\email{koltona@cab.cnea.gov.ar} \affiliation{Centro At\'omico Bariloche,
8400 S.C. de Bariloche, Argentina}
\author{Alberto Rosso} \email{rosso@lptms.u-psud.fr}
\affiliation{CNRS;
Univ. Paris-Sud, LPTMS UMR 8626, Orsay Cedex, F-91405, France}
\author{Thierry
Giamarchi}\email{Thierry.Giamarchi@unige.ch} \affiliation{Universit\'e
de Gen\`eve, DPMC, 24 Quai Ernest Ansermet, CH-1211 Gen\`eve 4,
Switzerland}
\author{Werner Krauth}
\email{krauth@lps.ens.fr}
\affiliation{CNRS-LPS\\ Ecole Normale Sup{\'{e}}rieure, 24 rue Lhomond,
75231 Paris Cedex 05, France}

\begin{abstract}
We study the steady state of driven elastic strings in disordered media
below the depinning threshold. In the low-temperature limit, for a fixed
sample, the steady state is dominated by a single configuration, which we
determine exactly from the transition pathways between metastable states.
We obtain the dynamical phase diagram in this limit.  At variance with a
thermodynamic phase transition, the depinning transition is not associated
with a divergent length scale of the steady state below threshold, but
only of the transient dynamics.  We discuss the distribution of barrier
heights, and check the validity of the dynamic phase diagram at small but
finite temperatures using Langevin simulations.  The phase diagram continues
to hold for broken statistical tilt symmetry. We point out the relevance
of our results for experiments of creep motion in elastic interfaces.

\end{abstract}
\maketitle

\section{Introduction}
\slabel{introduction}

Disordered elastic systems are ubiquitous
in nature.  They appear as interfaces such as
magnetic~\cite{lemerle_domainwall_creep,
yamanouchi_creep_ferromagnetic_semiconductor2,
repain_avalanches_magnetic,
metaxas_magneticwall}
or ferroelectric~\cite{paruch_ferro_roughness_dipolar,paruch_ferro_quench}
domain walls, contact
lines~\cite{moulinet_distribution_width_contact_line2},
fractures~\cite{ponson_fracture,alava_fracture},
or as periodic structures, such as vortex
lattices~\cite{blatter_vortex_review,du_aging_bragg_glass},
charge density waves~\cite{nattermann_cdw_review} and Wigner
crystals~\cite{giamarchi_electronic_crystals_review}.  They all share
the competition between elastic forces that tend to order the system
and the microscopic disorder that seeks to distort its structure. This
competition manifests itself in the static properties, leading to a
roughness of the interfaces or a distortion of periodic order.

In addition, disorder leads to pinning and thus also affects the dynamical
properties. Indeed  practically all the above systems may be displaced
through an external force (magnetic or electric field for magnetic
or ferroelectric domain walls, current for vortices, etc.). The motion
directly influences central observables of the system (e.g. magnetization
for magnetic domain walls, voltage for vortices etc.).

The static properties of disordered elastic systems are now well
understood. It has for example been established that interfaces (on which
we will focus in this paper) become rough in the presence of disorder.
The roughness is characterized by an exponent $\zeta\equil$, which depends
only on the universality class of the disorder, the dimension of the
interface and the nature of the elastic forces. The dynamical properties
are less well characterized. At zero temperature, disorder leads to the
existence of a critical pinning force $f_c$, the depinning threshold,
below which the interface is immobile, and above which steady-state motion
sets in. The dynamics at finite temperature is even more difficult to analyze than 
the zero-$T$ behavior.

It has been particularly fruitful to consider the depinning transition
as a regular critical phenomenon, with the velocity playing the role of
an order parameter~\cite{fisher_depinning_meanfield}. 
In this framework, the depinning transition appears
linked to the existence of a correlation length $\xi$ which diverges
at the depinning transition, and to the presence of critical
exponents, both for this length $\xi \sim (f-f_c)^{-\nu\dep}$ and for the
velocity $v \sim (f-f_c)^\beta$. For zero temperature, the analogy of
the dynamical depinning transition with equilibrium critical
phenomena  has been checked directly by numerical approaches and by
analytical techniques such the functional renormalization group.

How to extend the analogy with critical phenomena to finite
temperatures has not been completely evident.  Measurements of the
thermal rounding of the depinning transition, or observations of
scaling for transient dynamics indicate that these ideas carry over
to finite temperature \cite{chen_marchetti,vandembroucq,bustingorry_throunding}.
 However a direct study of the motion of such
systems is very difficult for forces below the depinning forces, at
finite temperature. Indeed in that case, the motion takes place by
thermal activation over barriers, leading to extremely long activation
times. This renders numerical techniques such as the molecular dynamics
inefficient.

At finite temperature, the dynamics of a disordered system is very
difficult to simulate, because of the high excitation barriers and the
presence of thermal noise. Standard dynamical algorithms become very
inefficient because the system is frozen in a local minimum. Advanced
dynamical simulation methods, such as the BKL algorithm~\cite{bortz_BKL}
and its variants cannot always be applied because the systems are
in fact equilibrated on some length scale, leading to a futility
problem~\cite{Pluchery,Krauth_smacbook}: Not only does it take a long time
to pass a barrier, but before doing so, the system will have undertaken
a very large number of (futile) moves among local configurations.

This problem can often be overcome in statics, where we need not follow
the very slow dynamics because observable averages are given by the
Boltzmann weight for each configuration. For this reason, many methods
(transfer matrix, optimization algorithms, advanced Monte Carlo methods)
allow one to characterize the thermodynamics of disordered elastic
systems with an effort polynomial in the system size.

Similarly, although analytical techniques
such as the functional renormalization group have been used with
success to investigate the motion at finite temperatures, they lead
to complicated equations that have been solved so far only
for the \quot{creep} regime of very small forces $f \to 0$. Moreover,
these techniques rely on an expansion around four spatial dimensions and
are thus not very well adapted to tackle quantitatively realistic one-
or two-dimensional interfaces.

In Ref \cite{kolton_depinning_zerot2} we introduced a novel numerical method 
which allows to follow the  motion of an interface at finite temperature,
without running into the above-mentioned difficulties. This algorithm
was used to demonstrate that the analogy between the depinning and critical
phenomena is incomplete. In fact, the steady-state motion lacks the
divergent length for $f\to  f_c$ from below. In the present paper,
we discuss the algorithm in detail. We use it to study the various
properties of an interface close to depinning. In addition to the
questions of the steady-state motion and the corresponding divergent
length scales and the roughness of the lines, we study
the distribution of activation barriers during the motion. This is
crucial since disordered elastic systems are glasses, with a priori
divergent barriers. We also compare the results of our algorithm
to molecular dynamics simulations, in order to check that taking the $T
\to 0$ limit before the thermodynamic limit does not introduce artifacts.

The outline of this article is as follows: After a short review, in
\sect{three_regimes}, of the statics and the zero-temperature dynamics of
disordered elastic systems, we outline in \sect{dynamic_phase_diagram}
the finite-temperature dynamical phase diagram obtained with our
method.  \sect{low_temp_dyn} discusses the special properties of
the low-temperature dynamics, which are used in the algorithm, and
\sect{numerical_results} contains a  detailed description of our numerical
results.  A general discussion, including the prospects for experiments in
\sect{discussion_conclusions}, concludes this paper. Technical details
of the algorithm and several mathematical proofs of key properties of
the low-temperature dynamics are contained in \app{algorithm}, whereas
\app{longrange_elasticity} resumes several properties of long-range
elastic systems.

\section{Basic notions}
\slabel{three_regimes}
In order to render our paper self-contained, we review in the present section
how the competition between disorder and elasticity manifests itself in the
statics and the zero temperature dynamics of elastic manifolds. 
We stress the difference between equilibrium properties and the 
non-equilibrium steady-state
behavior. Furthermore, we introduce the three \quot{reference states},
the corner-stones of our analysis of the dynamical phase diagram of
\sect{dynamic_phase_diagram}.

\subsection{Elastic manifolds}
\slabel{elastic_manifolds}

We consider a $d$-dimensional interface separating a $d+1$-dimensional
covering space into two regions. The interface is free of overhangs
or loops, and may thus be described by a uni-valued displacement field
$h(x)$. \fig{schema} shows an elastic string ($d=1$) in a two-dimensional
random medium, the case that we concentrate on in this paper.

\begin{figure}
\centerline{
\epsfclipon \epsfxsize=9.0cm \epsfbox{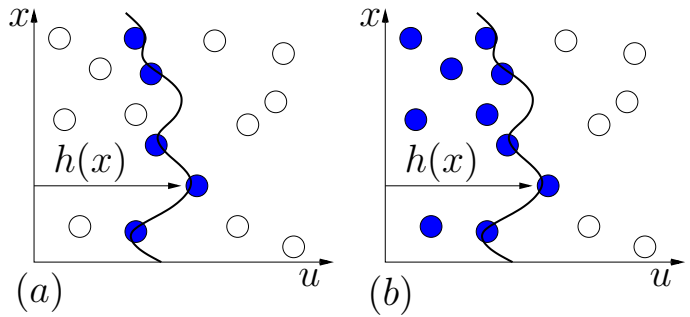} }
\caption{(Color online) An elastic string $h(x)$
in a random medium. (a) For random-bond disorder (RB),
its interaction with the impurities 
is local. (b) For random-field disorder (RF), the energy
$E_{\text{dis}}[h]$ depends on $V(u,x)$ for all
$u < h$.}
\flabel{schema}
\end{figure}

\begin{figure}
\centerline{
\epsfclipon \epsfxsize=9.0cm \epsfbox{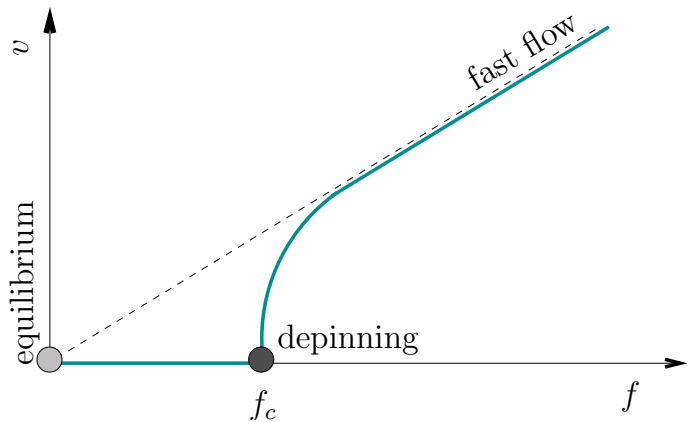} }
\caption{(Color on line) Velocity--force characteristics of an elastic manifold
at $T=0$. The three self-affine reference states are:
Equilibrium ($f=0$); Depinning ($f=f_c$); Fast flow ($f
\gg f_c$).}
\flabel{vvsf}
\end{figure}

The elastic energy $E_{\text{el}}[h]$ is minimal for the flat
manifold $h=\text{const}$. For short-ranged elastic interactions,
deviations from this configuration are often described by the harmonic energy
\begin{equation}
 E_{\text{el}}[h] = \frac{c}{2}\int d^dx \;(\partial_x h)^2
\elabel{elastic_energy}
\end{equation}
(with an elastic coefficient $c$). This case applies to magnetic and
ferroelectric domain walls as well as to vortex lattices and charge
density waves.

Dipolar forces \cite{nattermann_dipolar,paruch_ferro_roughness_dipolar,paruch_ferro_quench}, contact
lines for wetting \cite{joanny_contact_line}, and also crack propagation
\cite{gao_crack} represent the class of manifolds with long-range elastic
interactions. These systems are described by different harmonic forms
(see \app{longrange_elasticity}).

A second essential interaction is provided by the coupling of the manifold
to disorder. Two types of disorder correlations have been
much discussed in the literature: Random-bond (RB) disorder corresponds
to impurities that locally attract or repel the interface (as for the
wetting problem). In contrast, random-field (RF) disorder describes
pinning energies which are affected by the impurities inside the
entire region delimited by the interface (\fig{schema}). This situation
is encountered for example in magnetic systems when impurities
modify the local magnetic field. The disorder potential $E_\text{dis}$
generated by the impurities in these two cases is given by
\begin{equation}
   E_{\text{dis}}[h]       =     \int d ^d x\begin{cases}
    V(h(x),x) & \text{random bond}\\
    \ \int_0^{h(x)} d u V(u,x)       & \text{random field},
    \end{cases}
\elabel{disorder_class}
\end{equation}
where for each value of $x$, the function $V(u)$ is a
short-range-correlated Gaussian noise:
\begin{equation}
\overline{V(x,u)}=0,\;\;\;\overline{V(x,u)V(x',u')}=\delta^d(x-x') R(u-u')
\end{equation}
where the over-bar stays for the average 
over all disorder realizations and $R(u)$ is rapidly decaying function.

A third interaction appears in the presence of an external force.  With
strong intrinsic dissipation, the following over-damped zero-temperature
equation of motion applies:
\begin{equation}
\eta \partial_t h = -\frac{\delta (E_{\text{dis}}+E_{\text{el}})}{\delta h} + f =
      c \partial^2_x h + F_{\text{dis}}(h,x) + f,
\elabel{equation_of_motion}
\end{equation}
where $\eta$ is a friction coefficient and $F_{\text{dis}}$ is the
pinning force. For small external force $f$, the manifold ends up
pinned in a metastable state whereas it moves with finite velocity $v$
at larger force. The moving phase is separated from the pinned regime
by the critical force $f_c$~\cite{larkin_ovchinnikov_pinning}.

\subsection{Self-affine reference states ($T=0$) }
\slabel{self_afinne_ss}
\begin{table}\leftline{\small
\begin{tabular}{|c|c|c|c|c|c|c|}
\hline
 & $f=0$ & $f=0$ & $f=f_c$ & $f=f_c$  & $f \gg f_c$ & $f \gg f_c$  \\
 &  RB ~\cite{huse_exponent_line} & RF~\cite{fisher_functional_rg} & STS ~\cite{kolton_line_short_time} & no STS~\cite{kardhar_aniso_dep}
& EW &  KPZ ~\cite{kardar_parisi_zhang}\\
\hline
\hline
$\zeta$ & $2/3$ & $\sim 1$ & $\sim 1.25$ & $\sim 0.633$ & $1/2$ & $1/2$ \\
\hline
$z$ & $\infty$ & $\infty$ & $\sim 1.5$ & $1$ & $2$  & $3/2$  \\
\hline
$\nu$ & \eq{STS} & \eq{STS} & \eq{STS} & $\sim 1.733$ & $-$  & $-$ \\
\hline
\end{tabular}}\medskip
\caption{Zero-temperature critical exponents for an elastic string in the
three self-affine reference states (equilibrium, depinning, fast flow).}
\label{tabla_de_exponentes}
\end{table}

Three reference states are present in the velocity--force diagram at
zero temperature (\fig{vvsf}): the equilibrium (at $f=0$), the depinning
(at $f=f_c$), and the fast flow (at $f \gg f_c$). In these states,
the manifold is spatially self-affine. This means that lengths $x$ and
displacements $h$ above a small cut-off (set by the size of 
the corresponding Larkin domain~\cite{tanguy_larkin}) 
can be rescaled as $x' = b x$ and $h' = b^{\zeta} h$ into a new interface 
$h'(x')$ which is
statistically equivalent to $h(x)$. $\zeta$ is the state's characteristic
roughness exponent. Even at finite temperature, manifolds can be analyzed
in terms of these reference states, as we will discuss later.

\subsubsection{Equilibrium ($f=0$)}
\slabel{equilibrium_f_0}
The large-scale properties of elastic manifolds at equilibrium are
independent of temperature \cite{huse_exponent_line,giamarchi_vortex_long,
fisher_functional_rg,kardar_exponent_line,kardar_zeta_dp}, because
the disorder remains relevant. Indeed, sample-to-sample fluctuations
of the ground-state energy grow with the system size $L$ as $E_{\text{gs}}
\sim L^{\theta}$ with a positive exponent $\theta$.

The values of $\zeta\equil$ and $\theta$ depend on the dimension $d$, the range
of the elastic interactions and on the type of disorder (RB or RF), but
they are not independent: In a system of size $L$, the displacement $h$
scales as $L^{\zeta\equil}$, and the short-range elastic energy as $L^{2 \zeta\equil+d
-2}$ (see \eq{elastic_energy}). At equilibrium, the elastic energy and
the disorder contribution should scale in the same way. This implies
the scaling relation
\begin{equation}
\theta=2\zeta\equil+d-2.
\elabel{theta_relation}
\end{equation}
The motion is governed by the minimal energy
barriers between metastable configurations. The barrier between
configurations which differ over a size $l$ is of order $U \sim l^{\psi}$
for large $l$, with a positive barrier exponent $\psi$. This power
law is responsible for the logarithmically slow relaxation towards
equilibrium~\cite{fisher_random_scaling,kolton_relax}. The relation
$\psi=\theta$ is widely accepted~\cite{drossel_barrier} (see, however,
\cite{monthus_garel_barrier_exponent}, where $\psi=d/2$ is proposed).

\subsubsection{Depinning ($f = f_c$)}
\slabel{depinning}

At the depinning threshold $f=f_c^+$, the moving manifold is self-affine
both in space and in time. This means that displacements $h(t)- h(0)$
on scales $x$ and at time intervals $\sim t$ are statistically equivalent to
displacements on scales $a^{\zeta\dep/z\dep} x$ at times $\sim a t$ (here $z\dep$ is
the dynamic exponent). We note that in equilibrium the logarithmically
slow dynamics prevents self-affinity in time.

Above the depinning threshold ($f \gtrsim f_c$), the velocity vanishes
with a characteristic exponent, $v\sim(f-f_c)^\beta$, and the motion is
characterized by avalanches of a divergent typical size $\xi$,
\begin{equation}
\xi \sim (f - f_c)^{-\nu\dep}.
\elabel{correlation}
\end{equation}
In the language of critical phenomena, the velocity plays the
role of the order parameter and the force the role of the control
parameter~\cite{fisher_depinning_meanfield}.

The exponents $\beta$, $\zeta\dep$, $\nu\dep$, and $z\dep$
 are constrained by scaling
relations~\cite{ledoussal_frg_twoloops,nattermann_stepanow_depinning}. The
velocity of the manifold is related to the characteristic time of an
avalanche, $t \sim \xi^{z\dep}$, and to the distance the manifold advances
during this time $\xi^{\zeta\dep}$, as $v\sim \xi^{\zeta\dep-z\dep}$. This yields a
hyperscaling relation
\begin{equation}
 \beta=\nu\dep (z\dep-\zeta\dep).
\elabel{hyperscaling}
\end{equation}
Depinning exponents depend on the dimension $d$ of the manifold and on
the range of the elastic interactions but they are independent of the
type of disorder (RF or RB), merging the two equilibrium universality
classes into one~\cite{narayan_fisher_depinning,chauve_creep_long}.

Another scaling relation holds, both for depinning and equilibrium, if
the equation of motion preserves the statistical tilt symmetry
(STS)~\cite{narayan_fisher_depinning}. The tilt is a static force
$\epsilon(x)$ with vanishing spatial average. If one adds a tilt to
\eq{equation_of_motion}, and changes variables as $h(x,t)\rightarrow
h(x,t)+\nabla^{-2}\epsilon(x)$ or, in Fourier space,
$h(q,t)\rightarrow h(q,t) + q^{-2} \epsilon(q)$, the same equation of
motion is recovered with a new realization of the same disorder. The
statistical properties of the manifold are unchanged and the response
function behaves like in the pure system $\overline{\partial
h(q,t)/\partial \epsilon(q)} \sim q^{-2}$. Dimensional analysis of
\eq{correlation} implies that the force $\epsilon$ scales with the
exponent $-1/\nu$. On the other hand, self-affine displacements $h$
scale with the exponent $\zeta$, so that:
\begin{equation} 
\nu= \frac{1}{2-\zeta}.  \elabel{STS}
\end{equation}
Statistical tilt symmetry
symmetry is violated by a non-harmonic elastic energy or
by certain anisotropies of the random medium. This violation is relevant at depinning where a new universal behavior is observed
\cite{kardhar_aniso_dep,rosso_dep_exponent,wiese_aniso_dep_nosts}.
A phenomenological mapping has been proposed between a one-dimensional
string at the depinning transition and directed percolation
\cite{Tang_DPD,Buldyrev_DPD}. A concrete model belonging to this class
contains elastic energies stronger than harmonic, for example
\begin{equation}
 E_{el}[h] =\int d^d x \;\biggl[\frac{c}{2}(\partial_x h)^2 + c_4(\partial_x h)^4\biggr],
\label{eq:elastic_energy_anh}
\end{equation}
with $c_4>0$~\cite{rosso_dep_exponent}. We note that the violation of
statistical tilt symmetry changes the universality class at depinning,
but not in equilibrium.

\subsubsection{Fast flow ($f \gg f_c$)}
\slabel{fast_flow}
In the fast-flow reference state, for $f \gg f_c$, the quenched
 pinning force reduces
to an annealed stochastic noise~\cite{nattermann_stepanow_depinning} because 
in the co-moving frame, one has
$F_{\text{dis}}(h,x) = F_{\text{dis}}(\delta h+v t,x) \sim
F_{\text{dis}}(v t,x)$. For short-range correlated pinning force,
the strength of the disorder plays the role of an effective temperature
$T_{\text{eff}}$, since
\begin{multline*}
\overline{F_{\text{dis}}(v t,x)F_{\text{dis}}(v t',x')} \sim \Delta(0)\delta(v(t-t'))
\delta(x-x')  \\ = \frac{\Delta(0)}{v} \delta(t-t')\delta(x-x'), 
\end{multline*}
where $\Delta(0)$ measures the disorder strength. 
Since $d v/df \sim \eta^{-1}$ in this regime, 
it follows that $T_{\text{eff}} \sim \Delta(0)/(\eta v)$ from a generalized 
fluctuation--dissipation theorem. For the string, the fast-flow
state corresponds to the random walk.

Critical exponents for the three self-affine reference states
are summarized in Table~\ref{tabla_de_exponentes}.

\section{Dynamic phase diagram}
\slabel{dynamic_phase_diagram}

At non-zero temperature, the manifold moves with finite velocity for all
forces $f>0$, and the long-time dynamics reaches a steady state. 
Possible steady states are 
contained in a dynamical phase diagram exhibiting, at different length scales, 
the three reference states of \sect{three_regimes}. Above threshold, this diagram is well
understood in terms of the analogy with second-order phase transitions.

The steady-state dynamics below threshold can be studied by means of our
powerful algorithm which exploits the special properties of the dynamics
(see \sect{low_temp_dyn}). This algorithm yields a phase diagram which
differs from the standard picture of a second-order phase transition.

\subsection{Above threshold ($f > f_c$)}
\slabel{above_threshold}

\begin{figure}
\centerline{
\epsfclipon \epsfxsize=9.0cm \epsfbox{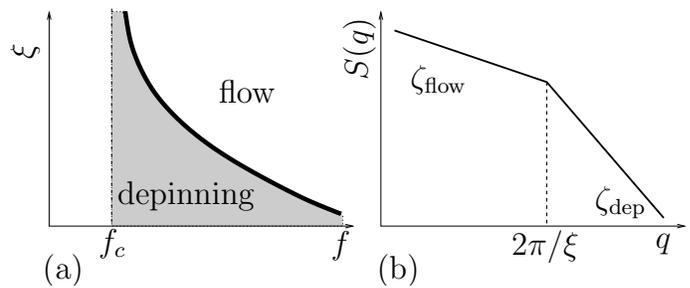} }
\caption{(a) Zero-temperature dynamical phase diagram.
(b) Schematic structure factor for $f > f_c$. The correlation length $\xi$ fixes the crossover 
between depinning and fast flow. }
\flabel{phase_diagram_Teq0}
\end{figure}

\begin{figure}
\centerline{
\epsfclipon \epsfxsize=9.0cm \epsfbox{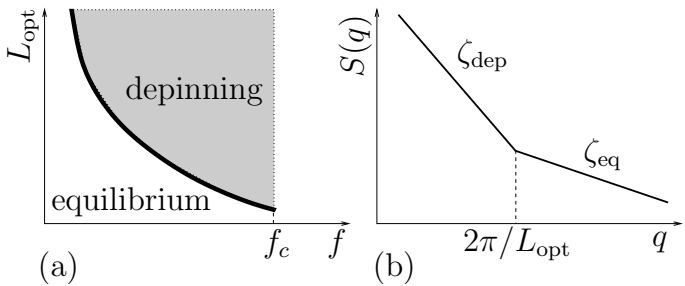} }
\caption{(a) Dynamical phase diagram for $f<f_c$ for vanishing temperature.
(b) 
Schematic structure factor for $f < f_c$. The length
$\Lopt$ fixes the crossover between equilibrium and depinning.}
\flabel{phase_diagram_Teq0plus}
\end{figure}

We first consider the region above the critical force where a
steady state exists even at zero temperature. Here the analogy
of the depinning transition with critical phenomena is well
understood~\cite{fisher_depinning_meanfield}: As in a thermodynamic
second-order phase transition, the connected two-point correlation
function of the order parameter is characterized by a correlation length
which diverges at the critical point. The steady-state velocity is the
order parameter of the depinning transition and its two-point correlation
function, 
\begin{equation}
\overline{\langle (v(x,t)-v)(v(0,t)-v) \rangle} \sim e^{-|x|/\xi},
\elabel{correlationdef}
\end{equation}
indeed diverges for $f \to f_c^+$. Here, brackets stand for the thermodynamic (or steady-state) average. In practice, this correlation function is not easily
accessible because the steady-state velocities necessitate the long-time
integration of the equation of motion \cite{duemmer}.

The correlation length $\xi$ separates two length scales in the
manifold (see \fig{phase_diagram_Teq0}a): on scales smaller than $\xi$,
the geometry of the interface is characterized by the exponents of
the depinning reference state (the critical phase in the language of
magnetic transitions). In contrast, on length scales larger than $\xi$,
the interface is governed by the exponents of the fast-flow reference
state (analogous to the ferromagnetic ordered phase of a magnetic
transition).  The length $\xi$ can be measured~\cite{duemmer}
through the structure factor~\cite{kolton_creep}, 
\begin{eqnarray}
S(q) &=& \overline{\biggl\langle \biggl|\frac{1}{L^{d/2}} \int d^d x \; h(x,t) \; e^{-i q x}\biggr|^2  \biggr\rangle} \nonumber \\
&=& \int d^d x\; e^{-i q x} \; \overline{\langle h(x,t) h(0,t) \rangle}, 
\elabel{S}
\end{eqnarray}
where the second equality makes use of spatial translation invariance.
For inverse lengths $q$ belonging to a self-affine regime with a single
roughness exponent $\zeta$, the structure factor takes the form
\begin{equation*}
 S(q) \sim q^{-(d+2\zeta)}.
\end{equation*}
The crossover between the depinning and the fast-flow regimes can be
conveniently extracted from the change of slope of the structure factor
$S(q)$ (see \fig{phase_diagram_Teq0}b).

\subsection{Below the depinning threshold}
\slabel{below_depinning}

Below the depinning threshold ($f < f_c$), at zero temperature, the
manifold is permanently pinned. We first discuss the phase diagram in the
limit of vanishing temperature, obtained with the methods of \sect{low_temp_dyn}.

As in the regime above threshold, the structure factor allows us to
access the self-affine regimes present in the interface. Our results from
\sect{numerical_results} are summarized in \fig{phase_diagram_Teq0plus}b.
A crossover length, $\Lopt$, associated to the maximal barrier encountered across the optimal path along the system, separates two roughness regimes:
On length scales smaller than  $\Lopt$ the
roughness of the interface is described by the equilibrium exponent
$\zeta_{\equil}$, corresponding to the paramagnetic phase in the
language of magnetic transitions. For distances bigger than $\Lopt$,
the roughness is described by depinning exponents (that is, the exponent
of the critical phase). This is at variance with standard critical
phenomenon, where the critical phase appears at large length scales only
at the critical point. Our results are summarized in the phase diagram of
\fig{phase_diagram_Teq0plus}a. Increasing the external force, 
$\Lopt$ decreases and when the depinning threshold is reached ($f \rightarrow
f_c^-$), $\Lopt$ coincides with the Larkin
length~\cite{tanguy_larkin}.

Our algorithm cannot 
access very small forces, but our results are compatible with predictions 
in the creep regime, that is, at low temperatures for $f \ll f_c$ 
\cite{ioffe_creep,nattermann_rfield_rbond,
chauve_creep_long, marcus_frgcreep, kolton_creep}. In this regime,
 a phenomenological scaling argument suggests that the
mean velocity is produced by activated jumps on a length scale which
diverges in the equilibrium limit $f \to 0$ as $\Lopt \sim
f^{-\nu_{\equil}}$. At this length, the typical energy barriers scale as
\begin{equation}
U(f) \sim \Lopt(f)^{\psi} \sim f^{-\mu},
\elabel{barrier_creep}
\end{equation}
where $\mu = \psi \nu_{\equil}$ is the creep exponent. The
velocity--force characteristics for small $f$ is thus a
stretched exponential. The functional renormalization group
\cite{chauve_creep_long} predicts  that on scales below
$\Lopt$ the system is in equilibrium and that scales larger
than $\Lopt$ are characterized by deterministic forward motion.

\begin{figure}
\centerline{
\epsfclipon \epsfxsize=9.0cm \epsfbox{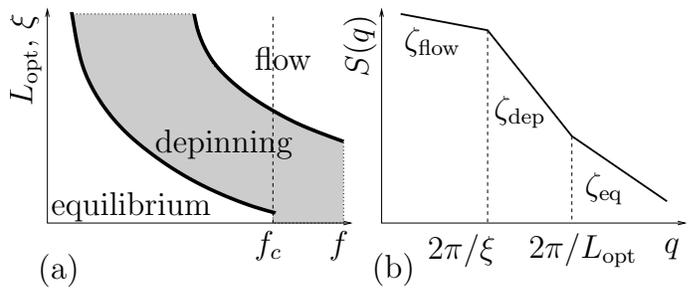} }
\caption{(a) Dynamical phase diagram at finite temperature.  (b)
Schematic structure factor below $f_c$.  The correlation length $\xi$
fixes the crossover between depinning and fast flow, and the length
$\Lopt$ the crossover between equilibrium and depinning.}
\flabel{phase_diagram_T}
\end{figure}
The structure factor at finite temperature is sketched in
\fig{phase_diagram_T}b. It presents three roughness regimes: for length
scales smaller than $\Lopt$, the interface is in equilibrium.
Between $\Lopt$ and $\xi$, it is characterized by depinning
exponents, and for larger lengths it is in the fast-flow regime. At
finite temperatures, the two crossover lengths $\Lopt$ and $\xi
\gg \Lopt$ diverge as $f \to 0$ (see \fig{phase_diagram_T}a).
While $\xi(f,T)$ diverges for $T\rightarrow 0$ below the depinning
threshold (see \fig{phase_diagram_Teq0}a), we find that
$\Lopt(f,T)$ saturates to a finite value $\Lopt(f,T=0)$.
The phase diagram of \fig{phase_diagram_T}a exhibits the connection 
between depinning and equilibrium.

\section{$T \to 0$ dynamics for finite samples}
\slabel{low_temp_dyn}

In this section, we discuss the detailed properties of the dynamics below
threshold in the the zero-temperature limit for a finite sample. We call
this the \quot{Arrhenius limit}, because the time $\Delta t$ to overcome
an energy barrier $U$ is governed by the Arrhenius formula $\Delta
t \sim \expl{U/T}$. We show in the present section that in this limit the steady-state dynamics is characterized by a
forward-moving sequence of metastable states of decreasing energy,
which we are able to compute using the exact algorithm described in
\app{algorithm}. Moreover, below threshold, for each value of
the external force, a unique dominant configuration is occupied with
probability one, in the same way as for a finite system at equilibrium the
occupation probability is entirely concentrated on the ground state.

\begin{figure}
   \centerline{
   \epsfclipon \epsfxsize=9.0cm \epsfbox{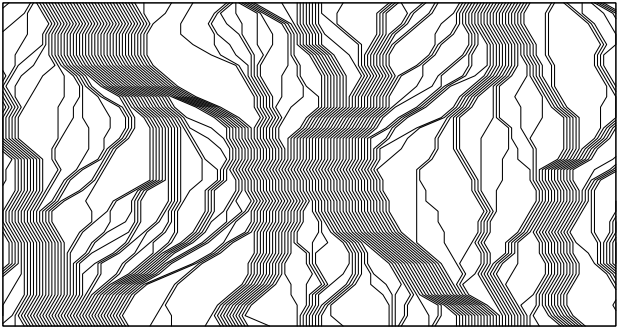} }
   \caption{ Sequence of metastable configurations $\SET{\dots,\alpha_k,
   \alpha_{k+1}, \dots}$ detected by our algorithm. For clarity,
   $\alpha_{k+1}$ is advanced by a small amount with respect to
   $\alpha_k$.  }
   \flabel{ordered_sequence}
   \end{figure}

\begin{figure}
\centerline{
\epsfclipon \epsfxsize=9.0cm \epsfbox{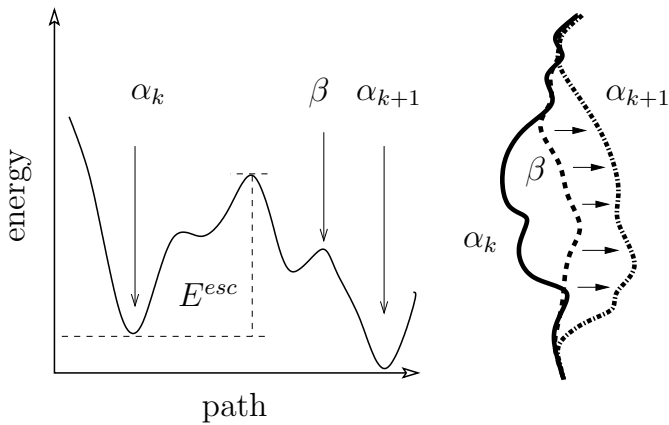} }
\caption{Escape path from a metastable configuration
$\alpha_k$ \emph{via} a configuration $\beta$ that relaxes towards a new
metastable configuration $\alpha_{k+1}$ of lower energy.
}
\flabel{optimal_path}
\end{figure}

\subsection{Ordered sequence of metastable states}
\slabel{ordered_sequence}

\begin{figure}
\centerline{
\epsfclipon \epsfxsize=7.0cm \epsfbox{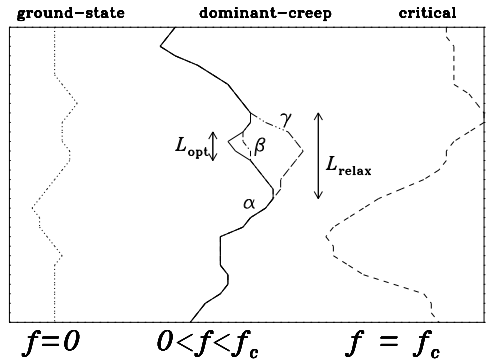} }
\caption{Ground-state, critical and dominant configurations below $f_c$.
An excitation of size $\Lopt$ escapes the system from the dominant
configuration $\alpha$.  The configuration $\beta$ relaxes to the
metastable state $\gamma$ \emph{via} an avalanche of size $\Lrelax$.}
\flabel{dominant_configurations}
\end{figure}

We first consider two metastable configurations $\alpha$ and $\gamma$
at a fixed force $f$. A path $P_{\alpha \rightarrow \gamma}$ connecting
$\alpha$ with $\gamma$ is a sequence of configurations labeled with
the time parameter $t$, as prescribed by the dynamics of the system,
for example a discrete rule on a lattice, or a continuum Langevin
prescription. The path barrier $P_{\alpha \rightarrow \gamma}$
between $\alpha $ and $\gamma$ is given by
\begin{equation}
B[P_{\alpha \rightarrow \gamma}]=\max_t E[\alpha(t)]-E[\alpha],
\end{equation}
and the barrier between the two metastable configurations, $B_{\alpha
\rightarrow \gamma}$, is defined as the minimal path barrier over
all paths connecting $\alpha$ with $\gamma$.  In the Arrhenius limit,
the escape time from a metastable state $\alpha$ is dominated by the
minimal barrier ($E^{\text{esc.}}[\alpha]$) connecting $\alpha$ to a
configuration with lower energy.
\begin{equation*}
   E^{\esc}[\alpha]=\min_{\gamma} B_{\alpha \rightarrow \gamma}, \quad 
   \text{with $E[\gamma] < E[\alpha]$}.
\end{equation*}

The previous properties suggest the definition of a coarse-grained
dynamics characterized by a sequence of metastable states,
$\alpha_0,\alpha_1, ...,\alpha_k$ of decreasing energy connected by
the minimal barriers $E^{\esc}$ (see \fig{ordered_sequence} and
\fig{optimal_path}). The transition from $\alpha_k$ to $\alpha_{k+1}$
is irreversible. This does not mean that from $\alpha_{k+1}$, the
microscopic dynamics cannot visit $\alpha_{k}$ again, but rather
that escape paths from $\alpha_{k+1}$ towards a new configuration
$\alpha_{k+2}$ with yet lower energy exist.
 In fact, as shown in \app{algorithm},
the escape path can be implemented entirely with forward-moving steps,
in the same way as for the zero-temperature motion relevant in the
depinning problem \cite{rosso_vmc_string,middleton_theorem}. The same
has been pointed out~\cite{Igloi_SDRG} for the Sinai model, where the
effective dynamics involves forward moves only.

The sequence of metastable states has two remarkable properties. First,
all configurations situated between configurations $\alpha_k$ and
$\alpha_{k+1}$ have smaller escape energies than the configuration
$\alpha_k$ itself. Second, if in the backward direction, no
metastable configuration exists which lowers the energy of $\alpha$,
then the coarse-grained dynamics starting from $\alpha$ is always
forward-moving.  These two properties (proved in \app{first_theorem}
and \app{second_theorem}) imply that the steady-state coarse-grained
dynamics is always forward-directed. Furthermore, the metastable
configuration characterized by the largest $E_{\text{esc}}$ belongs
to the sequence of configurations of the coarse-grained dynamics.
This is the dominant configuration which, as discussed above, is
occupied with probability one in the Arrhenius limit. Finally, the
steady state of a sample with periodic boundary conditions describes a
periodic trajectory of metastable configurations, which is independent
of the initial configuration. These properties are analogous of the
one-dimensional problem of a particle on a ring, which has been solved
exactly~\cite{derrida_hopping_particle,ledoussal_creep_1d}.

The details of the algorithm are given in \app{algorithm}.  It enumerates
a complete set of dynamically relevant configurations. For driven
manifolds, this approach is simpler than in general~\cite{Pluchery}
because the excited configurations differ from the metastable
configuration on a length scale $\Lopt$ which remains finite for $L \to
\infty$. In addition, we can restrict ourselves to forward moves.

In \fig{dominant_configurations} we compare the characteristic
configurations of the low temperature dynamics: the dominant
configuration, the ground-state, and the depinning critical
configuration. These three configurations are the dominant states
for $f=0$, $0<f<f_c$ and $f=f_c$, respectively. Our algorithm
accesses also the configurations $\beta$ and $\gamma$ sketched in
\fig{dominant_configurations}: configurations $\alpha$ and $\beta$
differ on a length $\Lopt$, and $\alpha $ and $\gamma$ differ on a length
$\Lrelax$. $\Lopt$ and $\Lrelax$ are the characteristic dynamical lengths
in the Arrhenius limit.

\section{Numerical results}
\slabel{numerical_results}

Our algorithm determines, for each sample, the dominant configuration,
the escape barrier $E_{\text{esc}}$, the size of the optimal thermal
excitation $\Lopt$, as well as the size of the deterministic avalanche
$\Lrelax$ (see \fig{dominant_configurations}). These results are also
compared with the results obtained by other methods at finite and at
zero temperature.

\subsection{Analysis of the dominant configuration}
\slabel{geometrical_properties}

\begin{figure}
\centerline{ \epsfxsize=9.0cm \epsfbox{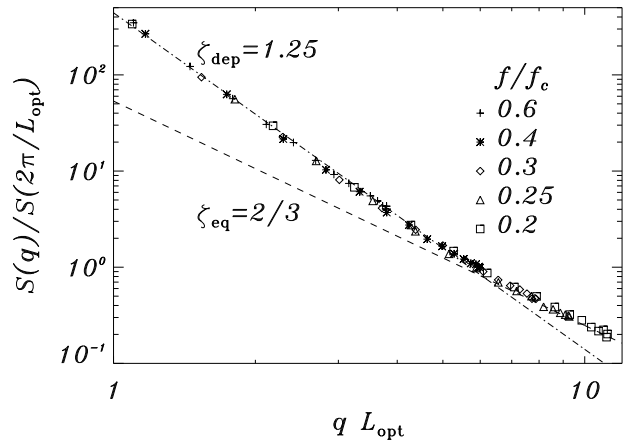}
}
\caption{Rescaled structure factor of the harmonic string in the
Arrhenius limit (averaged over $1000$ disorder realizations) for 
$L=16,32,64,128$. }
\flabel{lopt_is_both_geo_and_dyn}
\end{figure}

We first consider the random-bond elastic string with statistical tilt
symmetry. Complementing results of~\cite{kolton_depinning_zerot2}, we
show in  \fig{lopt_is_both_geo_and_dyn} the collapse of the structure
factor of the dominant configuration at different forces, with all
lengths rescaled by the disorder-averaged size of the thermal excitation,
$\Lopt$, which is obtained directly from the simulation. The remarkable
quality of the collapse (which is free of adjustable parameters) is also
due to the fact that the control parameter $f/f_c$, for each sample,
is obtained with the sample-dependent critical force.

In \fig{lopt_is_both_geo_and_dyn}, the change of regimes between
the small-$q$ region (governed by depinning) and the large $q$
regime (dominated by equilibrium) is manifest. We note that
$\Lopt$ characterizes the steady-state dynamics below $f_c$
in the Arrhenius limit in the same way as $\xi$ does above the depinning
threshold~\cite{fisher_depinning_meanfield,duemmer}.

We now consider the model with a hard metric constraint, which violates
STS. This case is particularly interesting because the
equilibrium roughness exponents are identical, whereas the depinning
exponents differ strongly. Moreover, in one dimension, the STS depinning
exponent is unphysical ($\zeta\dep \sim 1.25 >1$), and will probably not be
observable in nature.
\begin{figure}
\centerline{ \epsfclipon \epsfxsize=9.0cm \epsfbox{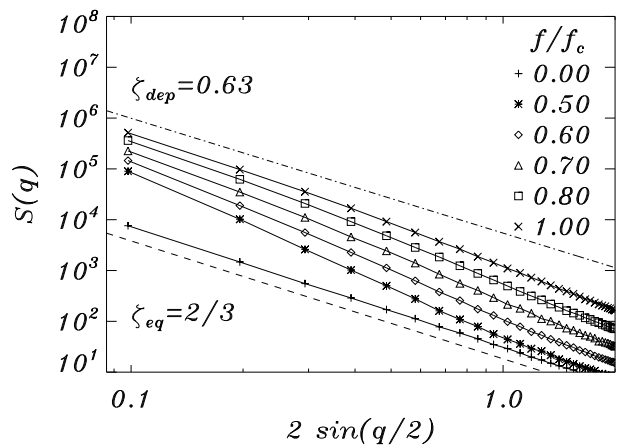} }
\caption{Steady-state structure factor for a string  with hard metric
constraint in the Arrhenius limit (averaged over 
$10000$ disorder realizations) for $L=M=64$. Curves for different
forces are shifted for clarity.}
\flabel{sofq_sos}
\end{figure}
\begin{figure}
\centerline{ \epsfclipon \epsfxsize=9.0cm \epsfbox{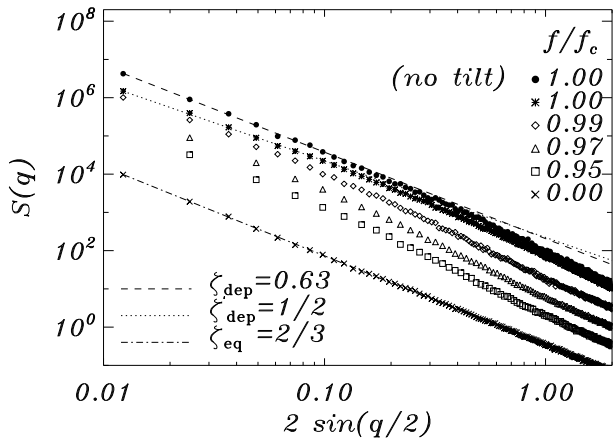} }
\caption{Steady-state structure factor for a tilted string  with hard metric
constraint in the Arrhenius limit (averaged over 
$1000$ disorder realizations) for $L=M=512, 256$. Curves for different
forces are shifted for clarity.}
\flabel{sofq_sos2}
\end{figure}
In \fig{sofq_sos}, we show the structure factor for the model with hard
metric constraints as a function of the force, from the statics, $f=0$,
to the depinning, $f=f_c$. The small-$q$ region is again consistent
with the depinning roughness exponent, and at large-$q$, we recover once
more the equilibrium behavior. However, the collapse of the curves for
different values of $f/f_c$ is not possible because the two exponents are now
very close in value. The crossover between the two regimes no longer depends on
the length scale $\Lopt$ alone, but also on the microscopic parameters
for which the string senses the presence of anharmonic corrections to
the elastic energy, that is, when $\overline{(h_{i+1}-h_i)^2} \sim 1$.

In the absence of STS, the roughness exponents at equilibrium and at
depinning are very close ($\zeta\dep \sim 0.63$ against
$\zeta\equi=2/3$).  However, the physics is very different in the two
regimes and reliable signals of the no-STS depinning regime are
present when a tilt is applied. A tilt can be realized through
a shift $s\cdot L$ imposed on the interface boundary conditions
(i.e. $h_L = h_0 + s \cdot L$ and $0<s<1$). Following Ref~\cite{goodman} the
roughness exponent for a tilted interface will differ from that of an
untilted one, and for a tilted string it is expected to be
$\zeta_{\text{tilt}}=0.5$~\cite{kardar_phys_rep}. 
In \fig{sofq_sos2} we show the structure
factor for the model with hard metric constraints in presence of a
tilt $s=0.5$. In the statics, $f=0$, the tilt has no effect. 
For $f>0$ we observe that tilted interfaces, at large length 
scales become less rough than untilted ones, in good agreement 
with $\zeta_{\text{tilt}}\approx 0.5$. This analysis confirms that 
in the creep regime the large scale structure of the string is 
described by deterministic processes belonging to the
corresponding depinning universality class of the system, 
regardless of the violation of the STS symmetry.


\subsection{Barriers}
\slabel{barriers}

\begin{figure}
\centerline{ \epsfclipon \epsfxsize=9.0cm \epsfbox{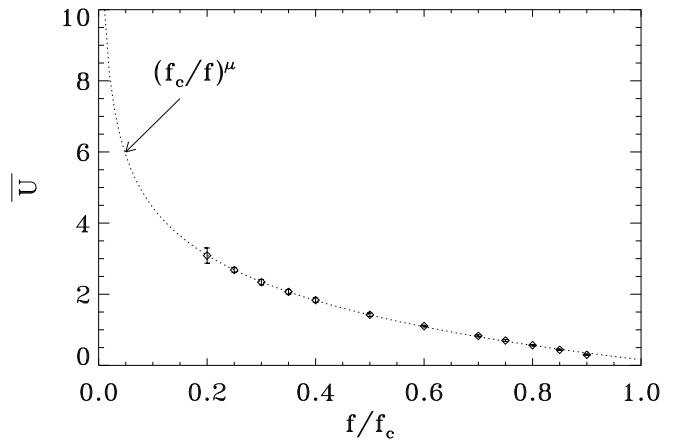} }
\caption{Mean escape barrier $U$ from the dominant configuration
for samples $L\times L^{\zeta\dep}$ with $L=32,64,128$ (the line is a guide to the eye). }
\flabel{barriers}
\end{figure}

\begin{figure}
\centerline{ \epsfclipon \epsfxsize=9.0cm
\epsfbox{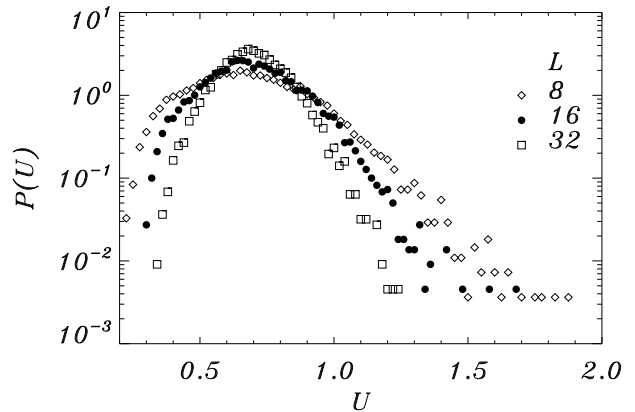}}
\caption{Distribution of $U$ for $11000$ disorder realizations (samples $L
\times L^{\zeta\dep}$ with $L=8,16,32$). The width of the distribution
decreases with increasing $L$.}
\flabel{barriersdist}
\end{figure}

In the Arrhenius limit, the velocity of the interface depends only on the
barrier height. Assuming a narrow distribution of barriers, we can
relate the mean velocity of the interface to the typical barrier $U(f)$:
\begin{equation}
v(f)=\Lopt^{\zeta\equil}  \left( \frac{\Lrelax}{\Lopt}
\right)^{\zeta\dep} e^{-\beta U(f)}
\end{equation}
In the creep regime, when $f$ is very small, we have $\Lrelax \sim
\Lopt$ and $U(f)$ is given by \eq{barrier_creep}. The phenomenological
expression for the velocity at low forces is the stretched exponential of the creep
formula~\cite{ioffe_creep,nattermann_rfield_rbond, chauve_creep_long, 
marcus_frgcreep,kolton_creep}
\begin{equation}
v(f) \sim \exp [-\beta (f_c/f)^{\mu}].
\elabel{phenomenological_formula}
\end{equation}
Our algorithm yields the escape barrier from the dominant configuration.
For a proper scaling~\cite{bolech_critical_force_distribution} of the 
sample dimensions $L \times L^{\zeta\dep}$, we may
identify this barrier with the characteristic barrier $U(f)$ and test
the phenomenological argument of \eq{phenomenological_formula}. In
\fig{barriers} we see that $U(f)$ increases with decreasing
$f$. This is consistent with the phenomenological arguments leading
to \eq{barrier_creep}, even if we cannot determine the exponent $\mu$,
because of the shortcomings of our algorithm at very small driving forces.

In \fig{barriersdist}, we show the barrier distribution as a function
of the size $L$ for samples scaled properly as $L \times L^{\zeta\dep}$.
The decay of the distribution for large $U$ appears faster than
exponential. Clearly, the distribution becomes narrower as the system
size is increased. Extrapolation of these results for an infinite system
leads to a well-defined value of $U(f)$, confirming the phenomenological
assumption of a typical barrier size.

\subsection{Finite temperature}
\slabel{finite_temperature}
\begin{figure}
\centerline{ \epsfclipon \epsfxsize=9.0cm \epsfbox{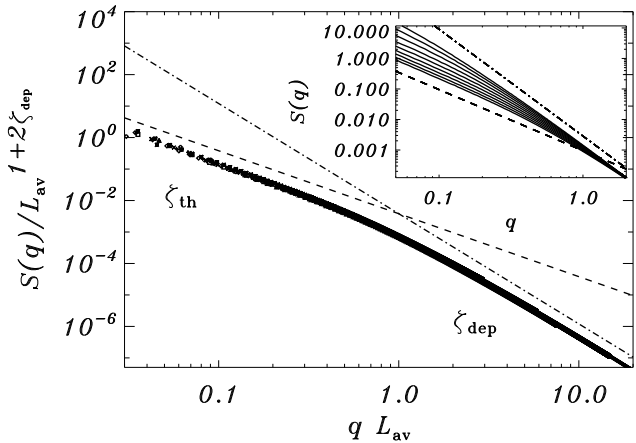} }
\caption{Finite-temperature Steady-state structure factor from Langevin 
simulations of an harmonic string of size $(L,M)=(1024,2048)$. 
The inset shows $S(q)$ for $T=0.05$ and
$f/f_c=0.8, 0.85, \dots 1.2$. 
The crossover length $\xi$ is a fit parameter 
for the collapse in the main panel.}
\flabel{langevin_steadystate}
\end{figure}

Our algorithm of \sect{low_temp_dyn} is powerful, but it takes the
$T \to 0$ limit before the thermodynamic limit. For a macroscopic
physical system, the order of these limits should be interchanged. To
confirm that the order of limits does not affect physical properties,
we have thus performed Langevin simulations of the equation of motion
\eq{equation_of_motion} at small finite temperatures with parameters
$\eta=c=1$ and with random-bond disorder from a normal distribution
interpolated with a cubic spline, as in \cite{rosso_depinning_simulation}.

In the inset of~\fig{langevin_steadystate}, we show $S(q)$ for the
steady-state motion of the string at $T=0.05$ for different forces around
$f_c$.  Two regimes are always present. They correspond to the depinning
at small scales and the flow regime at large scales. The crossover
between these regimes corresponds to the correlation length $\xi(f)$
of \fig{phase_diagram_T}. In the main panel of \fig{langevin_steadystate}
we have collapsed all data at small $q$ by using the scaling
$S(q)/S(2\pi/\xi) \sim G(q\xi)$ with a fitting parameter $\xi$.
\begin{figure}
   \centerline{ \epsfclipon \epsfxsize=9.0cm
   \epsfbox{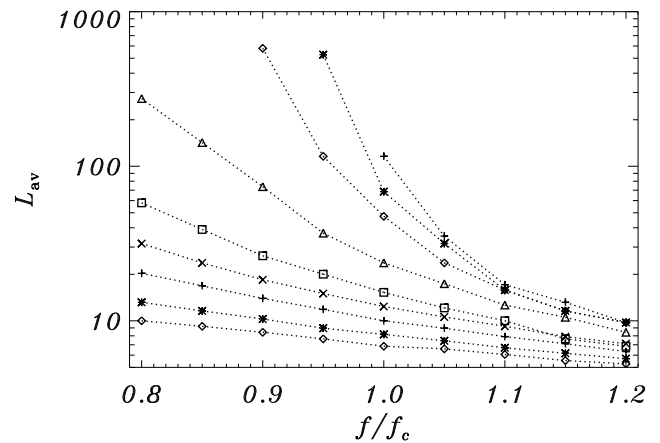} }
\caption{$\xi$ as a function of $f$ for different $T$ increasing from the
top to the bottom ($T = 0.0025,0.005,0.01,0.025,0.05,0.075,0.1,0.15,0.2$).}
\flabel{lavmd}
\end{figure}

The collapse of \fig{langevin_steadystate} provides the correlation
length $\xi$ as a function of force and temperature, and we show it
in \fig{lavmd} as a function of the reduced force $f/f_c$ for different
$T$. $\xi$ decreases monotonically with $f$, and tends to diverge near
$f_c$ in the $T \rightarrow 0$ limit, showing that $\xi$ is ultimately
controlled by the velocity of the string.  This behavior is fully
consistent with the schematic phase diagram of \fig{phase_diagram_T}.

In summary, the Langevin simulations confirm the scenario valid for
low-temperature Arrhenius dynamics for finite $T$. We find no indication
of a divergent (or even increasing) correlation length  as we approach
$f_c^-$.  The main modification at finite $T$ is that $\xi$ diverges at
$f=0$ instead of at $f_c$, since this divergence is controlled by the
vanishing of the steady-state velocity.

\subsection{Deterministic avalanches below $f_c$}
\slabel{deterministic_avalanches}

In the previous sections we considered two length scales below threshold,
namely the size $\Lopt$  of the optimal thermal excitation, and the
correlation length $\xi$ above which the disorder acts effectively as
a thermal-like noise.  Both lengths appear in steady-state quantities,
such as the structure factor $S(q)$ or the velocity $v$ of the interface.

Another length scale, $\Lrelax$ can be identified as the typical size of
the deterministic avalanche that drives the interface from the optimal
activated jump of size $\Lopt$ to the next metastable state.

$\Lrelax$ thus measures the distance between consecutive metastable
states. As shown in \fig{dynamical_lengths_vs_f}, $\Lrelax-\Lopt$
diverges approaching $f_c$ from below, with the characteristic exponent
$\nu\dep$.  However, $\Lrelax$ does not describe steady-state
properties below the depinning threshold, and represents no genuine
divergent length scale below the dynamic phase transition.

A length scale analogous to $\Lrelax$ appears also in the transient
dynamics at $T=0$ for an initially flat configuration which relaxes
up to the first pinned metastable state~\cite{hong_relaxtometa, kolton_line_short_time}.
  Here, a crossover separates the large length
scales keeping memory of the initial condition from the short length
scales characterized by the depinning roughness exponent. This length
scale diverges with the same exponent $\nu\dep$ as $\Lrelax$.
We can identify it with $\Lrelax-\Lopt$.

\begin{figure}
\centerline{ \epsfclipon \epsfxsize=9.0cm \epsfbox{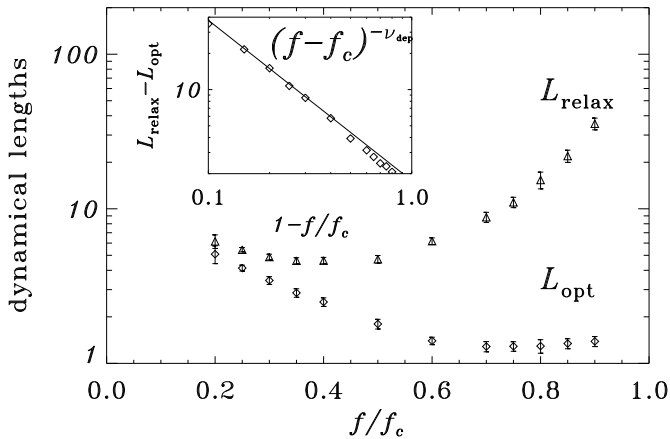} }
\caption{Behavior of the dynamical length $\Lrelax$. This length diverges
with the exponent $\nu\dep$ (see inset).}
\flabel{dynamical_lengths_vs_f}
\end{figure}

%
%

\section{Discussion and conclusions}
\slabel{discussion_conclusions}

We have studied in this paper the low-temperature dynamics of a driven
elastic string in a disordered medium below the depinning threshold.
In the (Arrhenius) limit of vanishing temperature, and at large scales,
the string behaves as at the critical force ($f=f_c$), and it shows the
universal properties of the depinning transition.  This result contradicts
the quasi-equilibrium picture of the creep motion, which assumes that
a small force simply moves the system from one metastable equilibrium
state to another, without changing the geometrical properties of these
states~\cite{ioffe_creep}. It also demonstrates that the analogy of
 the depinning transition with equilibrium critical phenomena is incomplete.
The effect of even a small force ($f \ll
f_c$) is in fact more radical: it drives the system away from the equilibrium
and into the depinning regime.  The equilibrium behavior of the string
is restricted to small scales, The crossover between the depinning and
the equilibrium regimes takes place at the length scale $\Lopt$ of the
optimal barrier. At the critical force, $\Lopt$ does not diverge,
and it equals the corresponding Larkin length.  Lowering the force, $\Lopt$
increases,  diverging in the limit of vanishing force.  Our findings
are compatible with the FRG predictions~\cite{chauve_creep_long}.

To arrive at these conclusions, we have used a powerful algorithm, which
is exact under the condition that the $T\to 0$ limit can be exchanged
with the thermodynamic limit. We have also checked its conclusions with
conventional Langevin simulations. At finite temperature, a much larger
length scale $\xi$ is introduced by the finite velocity of the system.
Length scales above $\xi$ are described in terms of the universal
fast-flow regime.

The approach of the critical force is not characterized by a
divergent length-scale of the steady-state properties.  Nevertheless,
a divergent length-scale $\Lrelax$ at $f_c$ describes the typical size
of the deterministic avalanches that are triggered by the activated
events. This deterministic part of the motion does not affect 
the steady-state geometry and can for example not be identified in
snapshot of the line's positions.

The distribution of barriers is narrow and the typical barrier grows
with decreasing the drive, in agreement with phenomenological
arguments. This shows that the motion is controlled by
the typical barriers, yielding a finite steady-state velocity,
rather than  by rare barriers described by extreme value
statistics~\cite{Monthus_zerocreepvelocity}.

We conjecture that our conclusions remain valid for $d$-dimensional manifolds
moving in $d+1$-dimensional space with short-range or long-range
elasticity. Moreover, the violation of the statistical tilt symmetry
changes only the large-scale geometry according to the change of
the universality class at the depinning transition. This may be
relevant for the interpretation of creep experiments in thin magnetic
films~\cite{lemerle_domainwall_creep,metaxas_magneticwall}, since systems violating STS
display a depinning roughness which is very close to
the equilibrium one. 

Our results may apply to different experiments probing the creep motion
of elastic interfaces.  We expect that an experimental verification of
our results is possible using, for instance, imaging techniques for
magnetic \cite{lemerle_domainwall_creep, repain_avalanches_magnetic,
metaxas_magneticwall} or electric \cite{paruch_ferro_roughness_dipolar,paruch_ferro_quench}
domain walls in thin films: $\Lopt$ could be extracted from the analysis
of a spatial correlation function, and $\Lrelax$ could be measured by
comparing consecutive (long-lived) metastable states when $f$ is
close to $f_c$, since then $\Lrelax$ controls the distances between
successive metastable states. Due to its transient nature $\Lrelax$
could be also measured by transient methods at $T=0$, where we relax
an uncorrelated initial condition until it locks to the first metastable
state at a given force.

Steady-state noise measurements (such as acoustic emission noise)
below the threshold could also provide an indirect verification
of our results. We expect the force dependent barriers shown
in~\fig{barriers} to control the waiting times between events
associated with deterministic avalanches of a diverging size
$\Lrelax$, translating into large noise peaks. This situation could
be realized in ferromagnets~\cite{sethna_noise_review} or in material
failure~\cite{bonamy_noisecracks,koivisto_paper}.

We thank P.~Le~Doussal and K.~J.~Wiese for illuminating discussions all
along this work.This work was supported in part by the Swiss NSF under MaNEP and Division II.

\appendix
\section{Algorithmic details}
\alabel{algorithm}

Our algorithm computes the dominant configuration of an elastic string
moving on a two-dimensional discrete $L\times M$ lattice with periodic
boundary conditions both in $L$ and in $M$. The line is
described by the variables $h(i)$, giving the displacement of the string
on the slice $i$, with $0\leq i < L$. The energy of the line is given by
\begin{equation}
E = \sum_i  \frac{1}{2} (h(i+1)- h(i))^2  - f h(i) + V(i,h(i)).
\label{eq:Energy_lattice}
\end{equation}
Periodic boundary conditions in $M$ are accounted for by the periodicity
of the disorder potential: $V(i,h)=V(i,h+M)$. Besides the harmonic
elastic energy (which preserves STS), we also consider a hard metric constraint:
\begin{equation}
|h(i)-h(i-1)| \le 1
\elabel{elastic_constraint}
\end{equation}
which violates STS.

We use elementary moves of the \quot{variant Monte Carlo} (VMC) algorithm
of Ref.~\cite{rosso_vmc_string,rosso_dep_exponent} which allows
for the simultaneous motion of $k+1$ adjacent sites by one lattice
spacing if no move of $k$ sites is energetically favorable. This
choice of dynamics avoids certain pathologies of the single-site
dynamics~\cite{rosso_vmc_string,rosso_dep_exponent}.

For each sample of the disorder potential, the equilibrium ground
state~\cite{huse_exponent_line, kardar_exponent_line}, the critical
depinning force $f_c$ and the associated zero-temperature configuration at
$f_c$ \cite{rosso_vmc_string,rosso_dep_exponent} can be computed easily.

Below threshold, two kinds of motions are present: the first is
deterministic, as defined by the VMC algorithm, and it relaxes each unstable
configuration toward a metastable state. The second is the activated
dynamics connecting $\alpha_k$ to $\alpha_{k+1}$ through the sequence
of VMC moves belonging to the optimal path, which is characterized by
the barrier $E_{\text{esc}}[\alpha_k]$.

\subsection{Complete-enumeration scheme}

We initialize the dynamics from the equilibrium ground-state, which by
its very nature cannot relax through backward moves when the external
force is positive. This allows us to restrict our attention to forward
moves only (see \app{second_theorem}).
At a given external force, we let the ground state relax toward the metastable
configuration $\alpha_0$. For each transition $\alpha_k \rightarrow
\alpha_{k+1}$ we build the archive of the visited configurations,
$\beta_i$, with increasing energy $E[\beta_1] \le E[\beta_2]\le E[\beta_3]
\le...$. A constant $E_{\text{cut}}$ is also introduced in order to
compute $E_\text{esc}[\alpha_k]$. The archive is initialized with a single
configuration, $\beta_1=\alpha_k$, and $E_{\text{cut}}$ is set to zero.
At each step the configuration $\beta_1$ is taken and erased from the
archive. Two operations are performed on this configuration. First, 
we update the maximal barrier
\begin{equation}
E_{\text{cut}}=\max(E_{\text{cut}},E[\beta_1]-E[\alpha_k])
\nonumber
\end{equation}
Second, we relax $\beta_1$ under the VMC dynamics to a configuration
$\beta'$. All configurations connected to $\beta'$ by VMC moves are
incorporated to the archive. If $\beta'$ has lower energy than $\alpha_k$,
the construction ends and the barrier $E_{\text{cut}}$ as well as the
last configuration erased from the archive $\beta_1$ and its associated
metastable state, $\beta'$ are output.

We can identify all the quantities defined in the previous sections:
\begin{equation}
\begin{array}{c}
\alpha_{k+1} = \beta' \\
E_{\text{esc}}[\alpha_k] = E_{\text{cut}} \\
\Lopt[\alpha_k] = \sum_i \Theta(h_{\beta_1}(i)-h_{\alpha_k}(i))\\
\Lrelax[\alpha_k] = \sum_i \Theta(h_{\beta'}(i)-h_{\alpha_k}(i))
\end{array}
\end{equation}
where $\Theta(x)$ is the step function with the prescription
$\Theta(0)=0$. Let us remark once again that for each realization of
disorder we analyze the data corresponding to the configuration with
the maximum value of $E_{\text{esc}}$. In \fig{orbit_metastable_states}
we sketch the output of our algorithm for a small sample
at different driving forces.
\begin{figure}
\centerline{
\epsfclipon \epsfxsize=10.0cm \epsfbox{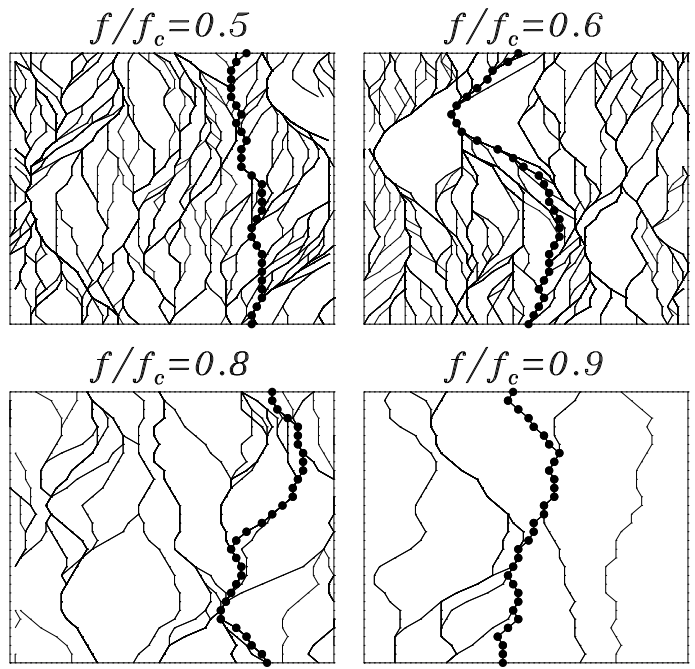} }
\caption{Metastable states of the low-temperature
in a $(L,M=32,64)$ system, for different 
forces $f<f_c$. The dominant configuration is emphasized.}
\flabel{orbit_metastable_states}
\end{figure}

\subsection{Bound on escape energies}
\alabel{first_theorem}
\begin{figure}
   \centerline{ \epsfclipon \epsfxsize=6.0cm \epsfbox{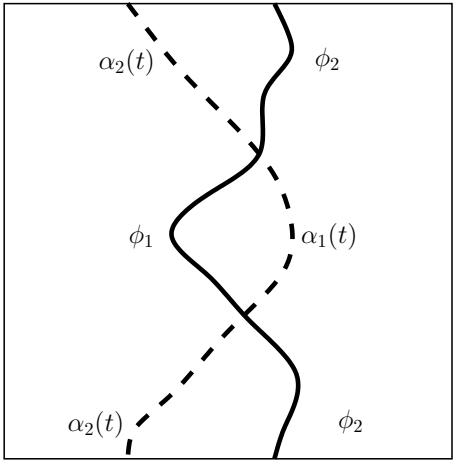} }
\caption{Decomposition of a 
configuration $\alpha(t)$ as a sum of
$\alpha_1(t)$ and $\alpha_2(t)$, the configuration $\phi$ by the sum
of $\phi_1$ and $\phi_2$, the configuration $\alpha_\phi(t)$ by the sum of
$\phi_1$ and $\alpha_2(t)$, and the  configuration $\phi_\alpha(t)$ by the sum of
$\alpha_1(t)$ and $\phi_2$.}
\flabel{firsttheorem}
\end{figure}

In this section, we show that all configurations between the metastable
configurations $\alpha_k$ and $\alpha_{k+1}$ have smaller escape
energies than the configuration $\alpha_k$ itself. This assures that the
coarse-grained dynamics must pass through the metastable configurations
with the largest barrier.

Let us consider a configuration $\alpha$ which moves to $\gamma$ in order
to pass through the minimal escape barrier $E^\esc_{\alpha}$. We show that
any configuration $\phi$ with $h_{\phi}(x) \ge h_{\alpha}(x)$ ($\forall
x$) and with an energy barrier $E^\esc_{\phi} \ge E^\esc_{\alpha}$
satisfies $h_{\phi}(x) \ge h_{\gamma}(x)$ ($\forall x$).

We suppose that the property is not true which means that $h_{\phi}(x)
\le h_{\gamma}(x)$ for some $x$. We follow the dynamical evolution of the
 line $\alpha(t)$ and compare it with the evolution
of two sets of configurations. The first one, $\alpha_{\phi}(t)$,
is defined as
\begin{equation}
h_{\alpha_{\phi}(t)}(x) = 
\begin{cases}
h_{\phi}(x) & \text{if $h_{\alpha(t)}(x)>h_{\phi}(x)$}  \\
h_{\alpha(t)}(x) & \text{otherwise}
\end{cases}.
\elabel{evolution1}
\end{equation}
The second one, $\phi_{\alpha}(t)$, is defined as
\begin{equation}
h_{\phi_{\alpha}(t)}(x)=
\begin{cases}
h_{\phi}(x) & \text{if $h_{\alpha(t)}(x) \le h_{\phi}(x)$}  \\
h_{\alpha(t)}(x)  &\text{otherwise}
\end{cases}.
\elabel{evolution2}
\end{equation}
We start by comparing $\alpha(t)$ and $\phi_{\alpha}(t)$. Because of the
metastability of $\phi$ we expect that, initially, $E[\alpha_\phi(t)]
\le E[\alpha(t)]$. However, this situation cannot continue up to
$t_{\text{end}}$ because, by definition, $\alpha(t)$ is the path which
encounters the smallest barrier. We call $\tilde{t}$ the smallest time
at which
\begin{equation}
 E[\alpha(\tilde{t})]< E[\alpha_{\phi}(\tilde{t})].
\elabel{cond1}
\end{equation}
Using the decomposition of \fig{firsttheorem}, for $t=\tilde{t}$, we
can write
\begin{multline*}
 E[\alpha(\tilde{t})]-E[\alpha_{\phi}(\tilde{t})]=\\
 E[\alpha_1(\tilde{t})] -E[\phi_1(\tilde{t})] +\Delta
 E_{\text{el}}[\alpha_{\phi}(\tilde{t})\rightarrow \alpha(\tilde{t}) ], 
\end{multline*}
where the variation of elastic energy is understood as
\begin{equation*}
 \Delta E_{\text{el}}[ \alpha_{\phi}(\tilde{t})\rightarrow
 \alpha(\tilde{t}) ] =
 E_{\text{el}}[\alpha_1(\tilde{t}),\alpha_2(\tilde{t})]-
 E_{\text{el}}[\phi_1(\tilde{t}),\alpha_2(\tilde{t})].
\end{equation*}
Let us now compare the energy associated to the configuration
$\phi_{\alpha}(\tilde{t})$ to the energy associated to the metastable
state $\phi$:
\begin{equation*}
 E[\phi_\alpha(\tilde{t})]-E[\phi] = E[\alpha_1(\tilde{t})]
 -E[\phi_1(\tilde{t})] +\Delta E_{\text{el}}[\phi\rightarrow
   \phi_\alpha(\tilde{t})].
\end{equation*}
The convexity of the elastic energy assures that
\begin{equation}
\Delta E_{\text{el}}[\phi\rightarrow \phi_\alpha(\tilde{t})] \le
\Delta E_{\text{el}}[ \alpha_{\phi}(\tilde{t})\rightarrow
\alpha(\tilde{t}) ].
\end{equation}
Using \eq{cond1} we conclude that $ E[\phi_\alpha(\tilde{t})]<
E[\phi]$. All paths connecting $\phi$ with $\phi_\alpha(\tilde{t})$
must overcome at least a barrier $E^{\text{esc}}_{\phi}$. This
means that there exists a time $t^*$ ($0<t^*<\tilde{t}$) for which $
E[\phi_\alpha(t^*)]- E[\phi] \ge E^{\text{esc}}_{\phi}$. We can use the
usual energy decomposition also for the configuration visited at the time
$t^*$ (see \fig{firsttheorem} for $t=t^*$). The following inequality
it is easy to prove:
\begin{equation}
E[\phi_\alpha(t^*)]- E[\phi] \le E[\alpha(t^*)]-E[\alpha_{\phi}(t^*)].
\end{equation}
This implies
\begin{equation}
E[\alpha(t^*)] \ge E[\alpha_{\phi}(t^*)]+ E^{\text{esc}}_{\phi}.
\end{equation}
On the other hand
\begin{equation}
E[\alpha] < E[\alpha_{\phi}(t^*)].
\end{equation}
It follows that
\begin{equation}
E[\alpha(t^*)]-E[\alpha] >  E^{\text{Esc.}}_{\phi}.
\end{equation}
This contradicts the initial assumption, demonstrating that
all configurations between the metastable
configurations $\alpha_k$ and $\alpha_{k+1}$ have smaller escape
energies than the configuration $\alpha_k$ itself.


\subsection{Effective forward motion}
\alabel{second_theorem}
In the present section we prove the following statement: If there is no
metastable configuration which lowers the energy of $\alpha$ in the
backward direction, the coarse-grained dynamics starting from $\alpha$
will be forever forward-directed. In other words, there exists an escape
path starting at $\alpha$ involving only forward motion. This allows us
to restrict the search of new configurations, and reduces the complexity
of our algorithm. It also allows us to discard the archive of accumulated
configurations whenever a new metastable minimum configuration is
encountered.

We suppose that the above statement is false by assuming that the
configuration $\gamma$ relaxes towards a configuration $\phi$ such that,
for some $x$, we have $h_{\phi}(x) < h_\gamma(x)$. Due to the convexity
of the elastic energy we may restrict ourselves to the region where the relaxation
is strictly backward and $h_{\phi}(x) \le h_\gamma(x)$ for all $x$. The
stability of $\alpha$ with respect to backward movements imposes
\begin{equation}
h_\alpha(x) \le h_{\phi}(x) \le h_\gamma(x) \;\; \forall x.
\elabel{condition1}
\end{equation}
The energies associated to these configurations satisfy
\begin{equation}
E[\alpha] > E[\gamma] > E[\phi].
\elabel{condition2}
\end{equation}
We now show that given \eq{condition1} and \eq{condition2}, there
exists a configuration $\phi_1$ for which
\begin{equation}
\begin{array}{cc}
h_{\phi}(x) \le h_{\phi_1}(x) \le h_\gamma(x) & \forall x\\ E[\alpha] <
E[\gamma] > E[\phi] >E[\phi_1] &
\end{array}.
\elabel{condition1b}
\end{equation}
In analogy with \app{first_theorem} we can compare the evolution
of $\alpha(t)$, $\alpha_{\phi}(t)$ and $\phi_\alpha(t)$ defined in
\eq{evolution1} and in \eq{evolution2}. The configuration $\phi_1$
corresponds to $\phi_\alpha(\tilde{t})$, where $\tilde{t}$ is the
smallest time at which
\begin{equation}
 E[\alpha(\tilde{t})]< E[\alpha_{\phi}(\tilde{t})].
\end{equation}
This construction can be applied to configurations $\phi_2,\phi_3,\ldots$
up to $\phi_n=\gamma$. In this case we have $E[\phi_n] =E[\gamma]$
and the statement is shown to be correct.

\section{Long-range elasticity}
\alabel{longrange_elasticity}
For interfaces with long-range interactions, the elastic energy can be
written in compact form in the harmonic approximation:
\begin{equation}
E_{\text{el}}=\int d ^d q |q|^\alpha h_q h_{-q}.
\elabel{long_energy}
\end{equation}
Here, the parameter $\alpha$ controls the range of the interactions.
The standard short-range interaction corresponds to $\alpha=2$.
The long-range interactions acting on the contact
line of a liquid meniscus ~\cite{joanny_contact_line} and on a
propagating crack front~\cite{gao_crack} yield $\alpha=1$. 

We expect the general scenario presented in this paper to remain valid
for the long-range case. The numerical values of the universal exponent
should depend on the range of the elastic interactions, parameterized
by $\alpha$.

The scaling relations discussed in this paper can be adjusted to the case
of general $\alpha$. From a dimensional analysis of \eq{long_energy}
we infer the generalization of \eq{theta_relation}
\begin{equation}
\theta=2\zeta_{\equil}+d-\alpha.
\elabel{theta_relation_long}
\end{equation}
The four exponents of the depinning transition
are still constrained by the hyperscaling relation
\eq{hyperscaling}. In presence of STS, the argument
given for $\alpha=2$ holds and the response function is given by
$\overline{\partial h(q,t)/\partial \epsilon(q)} \sim q^{-\alpha}$. The
STS scaling relation writes
\begin{equation}
\nu= \frac{1}{\alpha-\zeta}
\elabel{STS_long}
\end{equation}
A real-space representation of \eq{long_energy} involves fractional
derivatives ~\cite{zoia_fraclap}.

A discrete version of the force derived from \eq{long_energy} is given by
\cite{tanguy_discret,rosso_depinning_simulation}
\begin{equation}
f_{\text{el}}[h(i)]=\sum_{j \ne i} \frac{h(j)-h(i)}{|i-j|^{1+\alpha}}
\end{equation}
It presents strong finite-size effects for $\alpha \sim 2$ and all
choices $\alpha >2$ corresponds to the standard Laplacian force. It is
more convenient~\cite{zoia_fraclap} to use the following discretized force
\begin{equation}
f_{\text{el}}[h(i)]= \sum_{j \ne i} A(|i-j|)
  \left(h(j)-h(i) \right)
\end{equation}
with
\begin{equation}
A(|i-j|)= \frac{\Gamma(|i-j|-\frac{\alpha}{2}) \Gamma(\alpha+1)}
{\pi\Gamma(|i-j| +1+\frac{\alpha}{2})} \sin(\frac{\alpha}{2}\pi)
\end{equation}
where $\Gamma(x)$ is the Gamma function. This discretization holds for
all $\alpha >0$, and it is unaffected by slowing down for $\alpha\sim 2$.

\end{document}